\documentclass{iau}
\usepackage{graphicx}
\usepackage{multirow}

\title[] %% give here short title %%
{Astrophysics and Big Data:\\ Challenges, Methods, and Tools}

\author[Mauro Garofalo et al.] %% give here short author list %%
{Mauro Garofalo$^1$, Alessio Botta$^1$$^2$ \and Giorgio Ventre$^1$}

\affiliation{$^1$
DIETI, University of Naples Federico II, Via Claudio, 21 I-80125 Napoli, Italy
\\ email: {\{\tt mauro.garofalo, a.botta, giorgio\}}@unina.it \\
$^2$ NM2 srl, Napoli, Italy\\
[\affilskip]}

\pubyear{2016}
\volume{325} %% insert here IAU Symposium No.
\jname{Astroinformatics (AstroInfo16)}
\editors{A.C. Editor, B.D. Editor \& C.E. Editor, eds.}

\begin{document}

\maketitle

\begin{abstract}
Nowadays there is no field research which is not flooded with data. Among the sciences, Astrophysics has always been driven by the analysis of massive amounts of data. The development of new and more sophisticated observation facilities, both ground-based and spaceborne, has led data more and more complex (Variety), an exponential growth of both data Volume (i.e., in the order of petabytes), and Velocity in terms of production and transmission. Therefore, new and advanced processing solutions will be needed to process this huge amount of data. We investigate some of these solutions, based on machine learning models as well as tools and architectures for Big Data analysis that can be exploited in the astrophysical context.
%% add here a maximum of 10 keywords, to be taken form the file <Keywords.txt>
\end{abstract}

\firstsection % if your document starts with a section,
 % remove some space above using this command.
\section{Introduction}
In the last decades, the exponential growth of data has changed the way we do science. New and increasingly sophisticated astronomical facilities, both ground-based and spaceborne, produce massive amounts of data and they will be able to reach in few years a production rate in the order of Petabyte/Year. This data tsunami, both in terms of volume and velocity, will bring Astronomy in the big data era. In many situations where it could be infeasible to store all data produced by facilities, it is crucial to efficiently analyze these data and to produce a response close to real time. So, machine learning (ML) algorithms, combined with big data analytics (BDA) architectures, become necessary tools in knowledge data discovery process, helping to automatically recognize hidden patterns inside the data and to understand their relationships. In this paper, we provide an overview of BDA in Astrophysics describing the big data generated by the sky surveys, the ML algorithms and the BDA platforms that could help to gain more meaningful insight on these datasets.

\section{Big Data 3V in Astrophysics}
What is Big Data? According to \cite[Manyika et al. 2011]{Manyika_etal11} definition ``Big data is datasets whose size is beyond the ability of typical database software tools to capture, store, manage, and analyze''. This definition contains a time-variant aspect. Datasets that today can be considered as Big Data tomorrow could become ``normal'' data. Since this definition does not use any metric to define big data, we prefer to use \cite[Laney et al. 2001]{Laney01} definition that identifies the data growth challenge as three-dimensional, i.e., concerning an increase in Volume, Velocity, and Variety.\\
Following the latter definition, Table \ref{tab1} shows modern sky surveys in terms of the 3V characteristics.
\vspace{4mm}

\begin{table}[ht]
 \begin{center}
 \caption{Big Data 3V characteristics in astronomical sky surveys.}
 \label{tab1}
 \begin{tabular}{clrrr}
 & & \multicolumn{1}{l}{} & \multicolumn{1}{l}{} & \multicolumn{1}{l}{} \\
\textbf{Sky Survey} & & \multicolumn{1}{c}{\textbf{Volume}} & \multicolumn{1}{c}{\textbf{Velocity}} & \multicolumn{1}{c}{\textbf{Variety}} \\ \hline
\multicolumn{5}{c}{} \\
\begin{tabular}[c]{@{}c@{}}SDSS\\ \textit{Sloan Digital Sky Survey}\end{tabular} & & $~50$ TB & $~200$ GB per day & images, catalogs, redshits \\
\multicolumn{5}{c}{} \\
GAIA & & $~100$ TB & $~40$ GB per day & more then $100$ parameters \\
\multicolumn{5}{c}{} \\
\begin{tabular}[c]{@{}c@{}}Pan-STARRS\\ \textit{Panoramic Survey Telescope }\\\textit{and Rapid Response System}\end{tabular} & & $~5$ PB & $~5$ TB per day & images, catalogs \\
%\multicolumn{5}{c}{} \\
%EUCLID & & $~20$ PB & $~300$ GB per day & redshifts \\
\multicolumn{5}{c}{} \\
\begin{tabular}[c]{@{}c@{}}LSST\\ \textit{Large Synoptic Survey Telescope}\end{tabular} & & $~60$ PB & $~10$ TB per day & images, catalogs \\
\multicolumn{5}{c}{} \\
\begin{tabular}[c]{@{}c@{}}SKA\\ \textit{Square Kilometer Array}\end{tabular} & & $~3$ ZB & $~150$ TB per day & images, catalog, redshifts
\end{tabular}
\end{center}
\vspace{1mm}
 \scriptsize{
 {\it Notes:}\\ The column Volume refers to raw data produced at the end of the experiment. \\Values regarding Pan-STARRS, LSST, and SKA surveys refer to expected Volume and Velocity values.
 }
% \vspace{1mm}
\end{table}

\begin{itemize}
\item \textbf{Volume} refers to the amount of data. Different surveys produce datasets measurable in terabytes, petabytes, and even exabytes. This abundance forces to face challenges to capture, clean, transfer, store, analyze and visualize datasets.
\item \textbf{Velocity} refers to both the data generation rate and the processing time requirement. Depending on the speed of data arrival, they can be processed in batch, if they arrive at intervals of time, or streaming if they require a real-time analysis (e.g. during the cleaning phase). Table \ref{tab1} depicts surveys where the generation data is anticipated to exceed terabytes each night rate.
\item \textbf{Variety} refers to the data type, i.e., structured, semi-structured, unstructured, and mixed. Astronomical sky surveys could include images, redshifts, time series data, and simulation data. Data from various sources have their formats, which causes the challenge of integrating data, and they have hundreds of features, introducing the problem of high dimensional data visualization.
\end{itemize}

\section{Machine Learning Methods}
Machine learning was defined in 1959 by Arthur Samuel as ``the field of study that gives computers the ability to learn without being explicitly programmed''. ML allows to uncover hidden correlation patterns through an iterative learning by sample data (or past experiences) instead of being explicitly programmed. Common classes of problems that ML algorithms can solve are classification, regression, clustering, and outlier detection. These algorithms have been successfully used in astrophysics to solve different tasks. \cite[D'Istanto et al. 2016]{DIsanto_etal16}, presented an extensive investigation about classification performance of Random Forests, Multi-Layer Perceptron (MLP) with Quasi-Newton Algorithm, and K-Nearest Neighbors to classify transient objects, through experiments on the identification of cataclysmic variables, the separation between galactic and extra-galactic objects and identification of supernovae. \cite[Masters et al. 2015]{Masters_etal15} applied the Self-Organizing Maps (SOM) to the photometric redshifts problem, mapping the empirical distribution of galaxies in a multidimensional color space. A review of the use of data mining in astronomy was presented by \cite[Tagliaferri et al. (2003)]{Tagliaferri_etal03}. Table \ref{tab2} gives an overview on ML methods used to face some astrophysical problems.

\begin{table}[ht]
 \begin{center}
\caption{Machine Learning Algorithms for Astrophysics}
\label{tab2}
\begin{tabular}{cllll}
 & & \multicolumn{1}{c}{\textbf{Astrophysics Application}} & & \multicolumn{1}{c}{\textbf{Machine Learning Algorithms}} \\ \hline
\multicolumn{5}{c}{\textit{Supervised}} \\ \hline
Classification & & \begin{tabular}[c]{@{}l@{}}AGN Classification, \\ Globular Cluster Classification,\\ Photometric Classification\end{tabular} & & \begin{tabular}[c]{@{}l@{}}Neural Network, Genetic Algorithms, \\ Random Forest, Support Vector Machine,\\ Bayesian Network, K-Nearest Neighbors\end{tabular} \\
\multicolumn{1}{l}{} & & & & \\
Regression & & Photometric redshifts estimation & & \begin{tabular}[c]{@{}l@{}}Linear Regression, Random Forest, \\ Neural Network, K-Nearest Neighbors\end{tabular} \\
\multicolumn{1}{l}{} & & & & \\ \hline
\multicolumn{5}{c}{\textit{Unsupervised}} \\ \hline
Clustering & & Outlier Detection & & \begin{tabular}[c]{@{}l@{}}Principal Component Analysis, \\ K-Means, Self-organizing Map\end{tabular} \\
\multicolumn{1}{l}{} & & & & \\
\begin{tabular}[c]{@{}c@{}}Dimensional \\ Reduction\end{tabular} & & Parameter space reduction & & \begin{tabular}[c]{@{}l@{}}Principal Component Analysis,\\ Probabilistic Principal Surface\end{tabular}
\end{tabular}
 \end{center}
\vspace{2mm}
% \scriptsize{
 %{\it Notes:}\\
 %}
\end{table}

\section{A data mining tool for astrophysics: DAMEWARE}
DAMEWARE (Data Mining and Exploration Web Application REsource) is a web-based distributed platform for data mining with machine learning methods (\cite[Brescia et al. 2014]{Brescia_etal14}. The current release offers: as supervised methods four MLP implementations (i.e. classic Back Propagation, Genetic Algorithm, Quasi-Newton, and Levenberg-Marquardt Optimization Network), Random Forests, and Support Vector Machines (SVM); six SOM implementations, Principal Probabilistic Surfaces (PPS) and K-Means as unsupervised models. This set of techniques is useful for classification, regression, clustering, and feature selection tasks.
DAMEWARE aims at to be an invaluable machine learning toolkit to face astronomical tasks such as AGN classification, photometric redshifts prediction, and globular cluster classification.

\section{Big Data Analytics Platforms}
\begin{figure}[ht]
\vspace*{0.2 cm}
\begin{center}
 \includegraphics[width=3.5in]{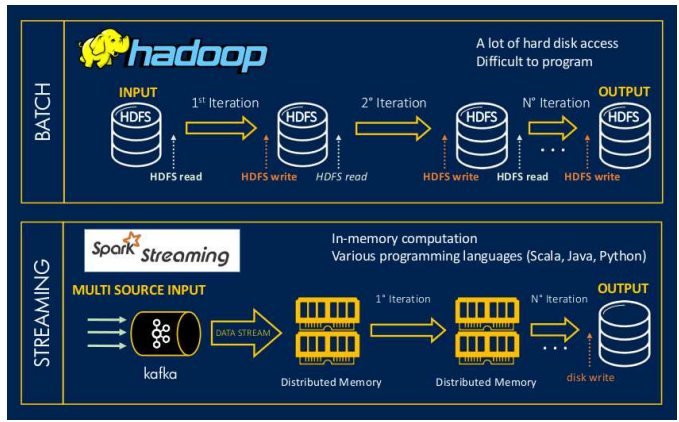}
 \caption{Hadoop vs Spark. Example of Big Data Analytics platforms for batch and streaming computing.}
 \label{fig1}
\end{center}
%\vspace*{1.0 cm}
\end{figure}

BDA architectures, deployed on Cloud or In-House Data Center, have become critical to face the computationally demand ML algorithms. \cite[Apache Hadoop]{HadoopLink} is an open source platform for distributed storage and batch processing of large data sets on clusters built from commodity hardware. Hadoop is widely adopted by many leading institutions for educational or production uses. Its services provide data access (HIVE and HBase), job scheduling (YARN), a distributed file system (HDFS), and data processing (Map-Reduce). \cite[Apache Spark]{Spark} is an open source BDA framework which makes it possible to analyze tons of data both in batch and streaming way. It provides two libraries for implementing machine learning: ML Lib, which provides ML models, and ML Pipelines handling the ML workflow, (i.e. data preparation, post-processing, and validations), helping to prepare and deploy the aforementioned models in a production environment. For both frameworks, Fig. \,\ref{fig1} shows a schematic representation of the architecture and the steps involved in the analytics process. Several vendors, such as Amazon, Google, and Microsoft, make available these frameworks as cloud services. \cite[Amazon Elastic MapReduce]{AmazonEMR} (EMR) provides an Hadoop cluster distributing the computation across multiple \cite[Amazon EC2]{AmazonEC2} instances. Moreover, EMR can run processing frameworks, such as Apache Spark and HBase, and interact with data in other \cite[Amazon Web Services]{AmazonAWS} (AWS) data stores. Google, with \cite[Cloud Dataproc]{Google}, provides Apache Spark services on Hadoop clusters for batch processing, querying, streaming, and machine learning. Finally, Microsoft, with \cite[Azure HDInsight]{Microsoft} provides an HDP Hadoop distribution in the cloud.

%(Fig.\,\ref{fig1})

\section{Conclusions}
The aim of this paper was to provide a brief overview of big data challenges in astrophysics. Astrophysical use cases which successfully exploit machine learning algorithms were presented. Some cloud implementations of BDA platforms have been cited as systems able to efficiently and quickly execute these computationally demanding algorithms. They will be interesting to face with astrophysics big data challenges, but currently, data transfer technologies are unsuitable for the amount of data involved in the computing. Thus, we keep following the ``move computing to the data'' paradigm.

\end{document}